\documentclass[twocolumn,showpacs,preprintnumbers,amsmath,amssymb]{revtex4}

\usepackage{graphicx}
\usepackage{dcolumn}
\usepackage{bm}

\begin{document}

\preprint{APS/123-QED}

\title{Role of Umklapp Processes in Conductivity of Doped Two-Leg Ladders}

\author{Patrick Byrne}
\affiliation{Department of Physics and Astronomy, McMaster University, Hamilton,
Ontario, Canada L8S-4M1}
\author{Eugene H. Kim}
\affiliation{Department of Physics and Astronomy, McMaster University, Hamilton,
Ontario, Canada L8S-4M1}
\author{Catherine Kallin}
\affiliation{Department of Physics and Astronomy, McMaster University, Hamilton,
Ontario, Canada L8S-4M1}

\date{\today}

\begin{abstract}
Recent conductivity measurements performed on the hole-doped two-leg ladder material $\mathrm{Sr_{14-x}Ca_xCu_{24}O_{41}}$ reveal an approximately linear 
power law regime in the $c$-axis DC resistivity as a function of temperature for $x=11$. In this work, we employ a bosonic model to argue that umklapp
processes are responsible for this feature and for the high spectral weight in the optical conductivity which occurs beyond the finite frequency Drude-like
peak. Including quenched disorder in our model allows us to reproduce experimental conductivity and resistivity curves over a wide range of energies. 
We also point out the differences between the effect of umklapp processes in a single chain and in the two-leg ladder.
\end{abstract}


\maketitle

\newcommand{\prp}{\phi_{\rho^+}}
\newcommand{\prm}{\phi_{\rho^-}}
\newcommand{\psp}{\phi_{\sigma^+}}
\newcommand{\psm}{\phi_{\sigma^-}}
\newcommand{\trp}{\theta_{\rho^+}}
\newcommand{\trm}{\theta_{\rho^-}}
\newcommand{\tsp}{\theta_{\sigma^+}}
\newcommand{\tsm}{\theta_{\sigma^-}}
\newcommand{\prpm}{\phi_{\rho^{\pm}}}
\newcommand{\pspm}{\phi_{\sigma^{\pm}}}
\newcommand{\trpm}{\theta_{\rho^{\pm}}}
\newcommand{\tspm}{\theta_{\sigma^{\pm}}}
\newcommand{\epi}[1]{e^{-\mathrm{i}\sqrt{\pi} #1}}
\newcommand{\efpi}[1]{e^{-\mathrm{i}\sqrt{4\pi} #1}}
\newcommand{\eppi}[1]{e^{\mathrm{i}\sqrt{\pi} #1}}
\newcommand{\epfpi}[1]{e^{\mathrm{i}\sqrt{4\pi} #1}}
\newcommand{\epmpi}[1]{e^{\pm \mathrm{i}\sqrt{\pi} #1}}
\newcommand{\emppi}[1]{e^{\mp \mathrm{i}\sqrt{\pi} #1}}
\newcommand{\xibpa}{\xi_{b+a}}
\newcommand{\xibma}{\xi_{b-a}}
\newcommand{\xibpma}{\xi_{b\pm a}}
\newcommand{\xisbpa}{\xi^*_{b+a}}
\newcommand{\xisbma}{\xi^*_{b-a}}
\newcommand{\xisbpma}{\xi^*_{b\pm a}}
\newcommand{\cpi}[1]{\cos\sqrt{\pi} (#1)}
\newcommand{\cfpi}[1]{\cos\sqrt{4\pi} (#1)}

\section{\label{sec:intro}Introduction}

Ladder materials have been the focus of much theoretical and experimental work 
over the last few years~\cite{rev}. These materials, which consist of planes of copper-oxygen
ladders separated by charge reservoir layers of other atoms, are interesting for 
a number of reasons. In terms of structure and behaviour there are similarities
to the high-$T_c$ cuprate materials. For example, hole doped two-leg
ladders exhibit a spin gap and the doped holes form pairs with a $d_{x^2-y^2}$
structure. Experimentally, a roughly linear temperature dependence of the resistivity has been observed in both doped ladders~\cite{osa}
and in the high-$T_c$ cuprates~\cite{gur}. In addition, the copper-oxygen ``ladders'' are weakly coupled to
one another, giving rise to a quasi one-dimensional system. This is interesting because 
there are analytical techniques for studying one-dimensional systems,
and, perhaps more importantly, one-dimensional electronic systems
exhibit Luttinger Liquid behaviour characterized by bosonic collective excitations and
spin-charge separation. Ladder systems provide an opportunity to study such behaviour, which
is not normally found in higher dimensional systems.

$\mathrm{Sr_{14-x}Ca_xCu_{24}O_{41}}$ is an example of an experimentally realizable ladder compound~\cite{mcc}.
This material, in comparison with other ladder materials, is easily fabricated and has 
received considerable attention from experimentalists. The structure of this compound consists of
three unique layers: a copper-oxygen two-leg ladder layer, a $\mathrm{CuO_2}$ chain layer, and a layer of
$\mathrm{Sr}$ and $\mathrm{Ca}$ atoms. By replacing $\mathrm{Sr}$ atoms with $\mathrm{Ca}$ atoms,
pre-existing holes can be moved from the chains to the ladders. For $x=0$, the ladders are undoped
and have a ground state with strong antiferromagnetic correlations.

Conductivity measurements have been performed on $\mathrm{Sr_{14-x}Ca_xCu_{24}O_{41}}$ for $x=8$ and $x=11$~\cite{osa}.
The $c$-axis (parallel to the legs) resistivity, $\rho_c$, and conductivity, $\sigma_{1c}$, for 
$x=11$ are of particular interest. For this doping the resistivity shows insulating behaviour at low temperatures
and metallic behaviour above $T\approx 100$ K. The conductivity has a low-frequency peak located
at a finite frequency ($\omega\approx 50$ cm$^{-1}$). At higher frequencies there is 
considerable spectral weight.

Conductivity of single one-dimensional chains in the presence of either quenched disorder or umklapp processes
has been studied in various theoretical papers~\cite{shu,gia1,gia2,gia3}. The effects of quenched disorder in two-leg ladders
have also been examined in detail~\cite{ori,kim}. In one such paper, Kim~\cite{kim} explains the
low temperature insulating behaviour and the low frequency optical conductivity peak of
the $\mathrm{Sr_{14-x}Ca_xCu_{24}O_{41}}$ two-leg ladder. 
Also, Qin et al.~\cite{qin} have recently examined the conductivity properties of the doped two-leg ladder
using a mean field holon-spinon treatment. However, an explanation of the metallic behaviour in
resistivity above $100$ K, specifically the monotonic increase in resistivity, for $x=11$ has not yet been presented. It has been suggested~\cite{kim}
that disorder, in the presence of pairing fluctuations, may provide an explanation provided spin fluctuations are suppressed
by the presence of a gap. The purpose of this paper
will be to examine the roles which quenched disorder and umklapp processes play in these regimes.
Specifically, we demonstrate that disorder cannot be responsible for the observed phenomena
whereas umklapp processes provide a plausible explanation, as they do for the high-$T_c$ cuprates~\cite{htc}. In addition, we will illustrate 
the differences between umklapp processes in a single chain and in a two-leg ladder. 

The plan of the paper is as follows. In Sec.~\ref{sec:model}, we present the model Hamiltonian 
used to describe the two-leg ladder with umklapp processes and disorder. This Hamiltonian may
be derived in the case of weak coupling, but becomes a phenomenological model in the strong coupling case.
In Sec.~\ref{sec:scale}, we present some scaling results which demonstrate that, at high energies, 
quenched disorder gives rise to a decreasing resistivity or a resistivity with little temperature
dependence (depending on whether spin fluctuations are present or not) whereas umklapp processes can give rise to a 
resistivity which is approximately linear in $T$.
In Sec.~\ref{sec:RG}, we present a renormalization group (RG) analysis for umklapp processes and disorder. 
Using this RG analysis and NMR results from Fujiyama et al.~\cite{fuj}, we argue that the scaling results of 
Sec.~\ref{sec:scale} apply to the $\mathrm{Sr_{3}Ca_{11}Cu_{24}O_{41}}$ two-leg ladder. We also discuss how perturbative calculations
of conductivity, which should be valid at energies well above the charge and spin gap energies, are 
also valid, at least qualitatively, until localization effects take over at low energies.
In Sec.~\ref{sec:pert}, we perform perturbative calculations for conductivity using the memory function formalism, in analogy
with Giamarchi's calculation~\cite{gia2} for the single chain with umklapp processes. Finally,
in Sec.~\ref{sec:curves}, we use the perturbative calculations of Sec.~\ref{sec:pert} along with the scaling equations of Sec.~\ref{sec:scale}
to produce conductivity curves which resemble the experimental results of Ref.~\cite{osa}.

\section{\label{sec:model}Model}

\subsection{Clean System}

The Hamiltonian for the clean two-leg Hubbard Ladder is given by
\begin{eqnarray}
H&=&-t\sum_{i,s} \left(c_{i+1,s}^\dagger c_{i,s}
+ c_{i,s}^\dagger c_{i+1,s}\right)\nonumber\\
&-&t\sum_{i,s} \left(d_{i+1,s}^\dagger d_{i,s}
+ d_{i,s}^\dagger d_{i+1,s}\right) \nonumber \\
&-&t_{\perp}\sum_{i,s} \left(c_{i,s}^\dagger d_{i,s}
+ d_{i,s}^\dagger c_{i,s}\right)\nonumber\\
&+&U\sum_i :n^c_{i,\uparrow} n^c_{i,\downarrow}:
+U\sum_i :n^d_{i,\uparrow} n^d_{i,\downarrow}:,
\end{eqnarray}
where the $c_{i,s}$'s annihilate fermions in spin state $s$ from the $i^\mathrm{th}$
site on one leg and the $d_{i,s}$'s perform the same operation on the other leg. Also, $:\ :$ represents normal ordering
with respect to the filled Fermi sea. To
investigate the properties of this model, we begin by making a transformation to the bonding/antibonding operators,
\begin{equation}
\left.\begin{array}{r}
b_{i,s}\\
a_{i,s}\end{array}\right\}=\frac{c_{i,s}\pm d_{i,s}}{\sqrt{2}}.
\end{equation}
The resulting free Hamiltonian ($U=0$) has a band structure consisting of two cosine dispersion relations separated by
$2t_\perp$, where $t_\perp$ is the hopping energy along a rung. 

Next we linearize about the Fermi points and take a continuum limit as in Ref.~\cite{emr}. The continuum Hamiltonian 
will be composed of chiral field operators, $\psi_{L/R,b/a,\uparrow/\downarrow}$, where $L/R$ refers to 
left/right moving fermions, $b/a$ refers to the bonding/antibonding band, and $\uparrow/\downarrow$ refers to 
spin up/down. Specifically, the interaction terms will be of the form $\mathrm{e}^{\mathrm{i}qx}\psi^\dagger_\alpha\psi^\dagger_\beta\psi_\delta\psi_\gamma$,
with $q$ being one of $\pm\{ 0, 2k_{b/a}, 4k_{b/a}, 2(k_a\pm k_b)\}$, where $k_{a/b}$ is the Fermi momentum of the antibonding/bonding band. At low energy 
scales, only the $q=0$ terms are important, unless one of the other $q$ values is near a reciprocal lattice vector (giving rise
to umklapp processes). For now we will ignore such processes and concentrate on the $q=0$ case. 

Since we are concerned with the low energy properties of the two-leg ladder, a cutoff of $O(1/a)$, where $a$
is the lattice spacing, may be introduced. This allows the well known Bosonization transformation~\cite{vond} to 
be applied to the continuum Hamiltonian. Following Lin et al.~\cite{lin}, we perform this procedure and ignore interaction terms
with irrelevant couplings. The resulting bosonic Hamiltonian is
\begin{eqnarray}
H=H_{\rho^+}+H_{\rho^-}+H_{\sigma^+}+H_{\sigma^-}+H_{\mathrm{int}},
\label{eq:ham1}
\end{eqnarray}
where 
\begin{eqnarray}
H_\nu = \frac{v_\nu}{2}\int dx \left[K_\nu\pi^2_\nu+\frac{1}{K_\nu}\left(\partial_x\phi_\nu\right)^2\right],
\end{eqnarray}
and
\begin{eqnarray}
H_{\mathrm{int}}&=&g_c\int \frac{dx}{a^2}\cos{\sqrt{4\pi}\theta_{\rho^-}}
\cos{\sqrt{2\pi}\phi_{b\sigma}}\cos{\sqrt{2\pi}\phi_{a\sigma}}\nonumber\\
&+&\int \frac{dx}{a^2}\left[g_{bs}\cos{\sqrt{8\pi}\phi_{b\sigma}}+
g_{as}\cos{\sqrt{8\pi}\phi_{a\sigma}}\right].
\label{eq:int}
\end{eqnarray}
A number of comments must be made about this Hamiltonian. First, we have neglected terms proportional to the difference between bonding
and antibonding Fermi velocities in order to eliminate cross terms from the free Hamiltonian. This is a good 
approximation if $t_\perp$ is smaller than $2t$ and neither band is very far from half filling. Second, in the non-interacting
limit, $H_{\mathrm{int}}\rightarrow 0$, $u_\nu$ becomes the Fermi velocity, and the Luttinger parameters, $\{K_\nu\}$, become unity.
Finally, something must be said about the boson fields themselves. Within the bosonization framework, fermionic
densities are represented by gradients of boson fields. In the present case, charge density in the bonding/antibonding band
is given by $\frac{1}{\sqrt{\pi}}\partial_x\phi_{b/a\rho}$ while spin density is given by $\frac{1}{\sqrt{\pi}}\partial_x\phi_{b/a\sigma}$.
The remaining fields in Eq.~(\ref{eq:ham1}) are defined as
\begin{eqnarray}
\phi_{\rho^\pm}&=&\frac{\phi_{b\rho}\pm\phi_{a\rho}}{\sqrt{2}}\nonumber\\
\phi_{\sigma^\pm}&=&\frac{\phi_{b\sigma}\pm\phi_{a\sigma}}{\sqrt{2}},
\end{eqnarray}
and
\begin{eqnarray}
\pi_i=-\partial_x\theta_i.
\end{eqnarray}

Lin et al.~\cite{lin} perform an RG analysis on the continuum fermion Hamiltonian for the two-leg ladder. By comparing the
couplings in Eq.~(\ref{eq:int}) to their counterparts in the fermion model, it is readily seen that $g_c$ and $g_{b/as}$
are relevant in the RG sense. This implies that the two spin modes are gapped and one charge mode is gapped. 
Also, in order to pin $\theta_{\rho^-}$, the two spin modes must first be pinned. That is,
the magnitude of the charge gap must be smaller than the spin gap. Although Eq.~(\ref{eq:int}) may be derived from the Hubbard for small $U$,
numerical work~\cite{noa} shows that there is no phase transition as one moves from weak to strong coupling. Therefore, we take this Hamiltonian
as a phenomenological model for the case of strong coupling as well.

The free part of the Hamiltonian in Eq.~(\ref{eq:ham1}) is written in terms of the $\phi_{\sigma^\pm}$ fields  
to eliminate cross terms which would appear if it had been written in terms of the $\phi_{b/a\sigma}$ fields. The interaction
part of the Hamiltonian is written in terms of the $b/a\sigma$ fields in order to elucidate the gap properties
of the system. However, it is convenient to express the entire Hamiltonian in terms of the $\sigma^\pm$ fields
when describing scaling and renormalization properties. The interaction part then becomes 
\begin{eqnarray}
H_{\mathrm{int}}&=&\frac{g_1}{a^2}\int dx\, \cos{\sqrt{4\pi}\theta_{\rho^-}}\cos \sqrt{4\pi}\phi_{\sigma^+}\nonumber\\
&+&\frac{g_2}{a^2}\int dx\, \cos{\sqrt{4\pi}\theta_{\rho^-}}\cos \sqrt{4\pi}\phi_{\sigma^-}\nonumber\\
&+&\frac{g_+}{a^2}\int dx\, \cos \sqrt{4\pi}\phi_{\sigma^+}\cos \sqrt{4\pi}\phi_{\sigma^-}\nonumber\\
&+&\frac{g_-}{a^2}\int dx\, \sin \sqrt{4\pi}\phi_{\sigma^+}\sin \sqrt{4\pi}\phi_{\sigma^-},
\label{eq:newform}
\end{eqnarray} 
where $g_1$ and $g_2$ are equal to $g_c$, but will flow differently under RG, and $g_{\pm}=(g_{as}\pm g_{bs})/2$.

\subsection{Scattering Mechanisms}

\subsubsection{Umklapp Processes}

In the previous section, interaction terms of the form $\mathrm{e}^{\mathrm{i}qx}\psi^\dagger_\alpha\psi^\dagger_\beta\psi_\delta\psi_\gamma$ were
ignored for non-zero $q$ values. However, if any of the possible $q$ values are close to a reciprocal lattice vector, then such terms will
become important. With the restrictions on $t_\perp$ and filling ($t_\perp$ small compared to $t$, and both bands not far from half-filling), the only 
$q$ values which can satisfy this condition are $\pm\{ 4k_{b/a}, 2(k_a+k_b)\}$. In these cases the net momentum transfers will be $2\pi /a-q$. 
When Bosonized, the umklapp terms become
\begin{eqnarray}
H_\mathrm{um}&=& \int \frac{dx}{a^2}\, \cos{(\sqrt{4\pi}\phi_{\rho^+}+\delta_{b+a}x)}\left[g_{u1}\cos\sqrt{4\pi}\theta_{\rho^-}\right. \nonumber\\
&-&\left. g_{u2}\cos\sqrt{4\pi}\phi_{\sigma^-}-g_{u3}\cos\sqrt{4\pi}\theta_{\sigma^-}\right]\nonumber\\
&+&g_{u4}\int \frac{dx}{a^2}\, \cos\left[\sqrt{4\pi}\left(\phi_{\rho^+}+\phi_{\rho^-}\right)+\delta_ax\right]\nonumber\\
&+&g_{u5}\int \frac{dx}{a^2}\, \cos\left[\sqrt{4\pi}\left(\phi_{\rho^+}-\phi_{\rho^-}\right)+\delta_bx\right],
\label{eq:umklapp}
\end{eqnarray}
where
\begin{eqnarray}
\delta_{b+a}&=& \frac{2\pi}{a}-2(k_a+k_b),
\end{eqnarray}
and
\begin{eqnarray}
\delta_{b/a}&=& \frac{2\pi}{a}-4k_{b/a}.
\end{eqnarray}

Eq.~(\ref{eq:umklapp}) reveals an important difference between umklapp processes on a single chain and on a two-leg ladder. For the chain, spin was completely
decoupled from the charge carrying mode~\cite{gia2}. For the ladder, umklapp processes couple the charge mode, $\rho^+$, to spin. The consequences of this coupling will 
be discussed later.

\subsubsection{Disorder}

For one dimensional conductors, disorder has typically been modeled by a random on-site potential~\cite{abr,gia1,ori},
\begin{eqnarray}
H_\mathrm{dis}&=& \sum_{i,p,s}\epsilon_{i,p}\,:n_{i,s,p}:
\end{eqnarray}
where $p$ is the leg label, $i$ is the site label and $s$ is the spin state label.
Also, the disorder strength, $\epsilon_{i,p}$, is taken from a Gaussian
distribution of width, $D$, which should be proportional to $v_F/\tau$, where $v_F$ is the Fermi velocity and
$\tau$ is the mean free path.

After switching to bonding/antibonding operators, taking the continuum limit and Bosonizing, the disorder Hamiltonian becomes
\begin{eqnarray}
H_\mathrm{dis}&=&\int \frac{dx}{a^2}\,\xi_1\epi{(\prp+\prm)}\cpi{\psp+\psm}\nonumber\\
&+&\int \frac{dx}{a^2}\,\xi_2\epi{(\prp-\prm)}\cpi{\psp-\psm}\nonumber\\
&+&\int \frac{dx}{a^2}\, \xi_3\left(\epi{(\prp-\trm)}\cpi{\psp-\tsm}\right.\nonumber\\
&+&\left.\epi{(\prp+\trm)}\cpi{\psp+\tsm}\right)\nonumber\\
&+&\int \frac{dx}{a^2}\, \xi_4\left(\epi{(\prm+\trm)}\cpi{\psm+\tsm}\right.\nonumber\\
&+&\left.\epi{(\prm-\trm)}\cpi{\psm-\tsm}\right)\nonumber\\
&+&H.c.,
\label{eq:bosdis}
\end{eqnarray}
where the $\xi$'s are taken from independent Gaussian distributions. Specifically, the disorder fields satisfy
\begin{eqnarray}
\left<\left<\xi_i(x)\xi^*_j(x^\prime)\right>\right>&=&aD_i\delta_{i,j}\delta(x-x^\prime),\nonumber\\
\left<\left<\xi\xi\right>\right>&=&0
\end{eqnarray}
where $\left<\left<\dots\right>\right>$ denotes disorder averaging.
Terms which produce scatterings from a Fermi point to itself have been ignored, as they are unimportant for conductivity calculations and may be 
eliminated by simple transformations~\cite{ori}.

Eq.~(\ref{eq:bosdis}) is useful for performing perturbative calculations above the spin and charge gaps. However, below these energies, simplified
forms for the Hamiltonian may be derived by integrating over quantum fluctuations which arise due to the pinning of a field and the rapid oscillations this
induces in the corresponding conjugate field. This procedure is explained in Appendix C of Ref.~\cite{ori}.

Some earlier experimental evidence~\cite{mag} points to a spin gap of about $300$ K for $\mathrm{Sr_{14-x}Ca_xCu_{24}O_{41}}$, with $x=11.5$. One possible explanation of the nearly linear power law
dependence of $\rho_c$ on $T$~\cite{kim}, as mentioned in the introduction, involves quenched disorder in the presence of pairing fluctuations without spin fluctuations (i.e. $\trm$ is free to fluctuate,
but $\psp$ and $\psm$ are pinned). If such a regime exists, the effective Hamiltonian may be derived by integrating over the fluctuations in $\tsp$ and $\tsm$.
The result is
\begin{eqnarray}
H&=&H_{\rho^+}+H_{\rho^-}+g_c\int \frac{dx}{a^2}\, \cos\sqrt{4\pi}\trm \nonumber\\
&+&\int \frac{dx}{a^2}\, \left[\xi_a\epi{(\prp-\prm)}\right.\nonumber\\
&+&\left.\xi_b\epi{(\prp+\prm)}+H.c.\right],
\label{eq:spingapham}
\end{eqnarray}
where terms which would contribute to physical properties by an amount of $O(D^2)$ have been dropped.

\section{\label{sec:scale}Conductivity and Scaling}

The real part of the conductivity is given by 
\begin{eqnarray}
\Re\sigma =\frac{-1}{\omega}\Im\left[\Pi(\omega,T)\right]
\label{eq:realcond}
\end{eqnarray}
where $\Pi$ is the current-current correlation function which may be calculated using the Kubo formula. In one dimension we have 
\begin{eqnarray}
\Pi(\omega) = -\mathrm{i}\int_{-\infty}^\infty dx \int_0^\infty dt\, e^{\mathrm{i}\omega t} \left<\left[j^\dagger(x,t),j(0,0)\right]\right>,
\label{eq:kubo}
\end{eqnarray}
where $j$ is the current operator, which can be readily obtained from the charge density operator, $\rho$, using the continuity equation, $\partial_t\rho
-\partial_xj=0$. For the present system, $\rho$ is given by
\begin{eqnarray}
\rho=\frac{2}{\sqrt{\pi}}\partial_x\prp,
\end{eqnarray}
yielding
\begin{eqnarray}
j=\frac{2}{\sqrt{\pi}}\partial_t\prp.
\label{eq:currentop}
\end{eqnarray}
Inserting this into Eq.~(\ref{eq:kubo}) gives
\begin{eqnarray}
\Pi(\omega,T)=\frac{4\omega^2}{\pi}G_{\mathrm{ret}}(\omega,T)
\end{eqnarray}
where
\begin{eqnarray}
G_{\mathrm{ret}}=\int_{-\infty}^\infty dx \int_0^\infty dt\, e^{\mathrm{i}\omega t} \left<\left[\prp(x,t),\prp(0,0)\right]\right>.
\end{eqnarray}

At high energies, the retarded Green's function, $G_{\mathrm{ret}}$, may be calculated using perturbation theory. However, if low energy properties are of
interest, then another approach must be taken. In this work, RG and scaling techniques~\cite{car} are used to obtain low energy information from
perturbative calculations of higher energy properties. Suppose energy is scaled in the following way: $E \rightarrow bE\ (b=e^{dl}$ 
and $E= \omega$ or $T)$, then the Green's function transforms as
\begin{eqnarray}
G_{\mathrm{ret}}(E,\{g\})=b^2Z_{\rho^+}G_{\mathrm{ret}}(bE,\{g(dl)\})
\label{eq:greenscale}
\end{eqnarray}
where $Z_{\rho^+}$ is due to ``wave function renormalization'', $\{g\}$
represents the set of coupling constants, and $\{g(dl)\}$ represents the set of coupling constants after scaling. Letting $Z_{\rho^+}=b^{-\gamma}$ and
iterating Eq.~(\ref{eq:greenscale}) gives
\begin{eqnarray}
G_{\mathrm{ret}}(E,\{g\})&=&\exp\left[\int_0^{l^*}dl\,(2-\gamma(l))\right]\nonumber\\
&\times&G_{\mathrm{ret}}(e^{l^*}E,\{g(l^*)\}).
\label{eq:greensrelation}
\end{eqnarray}
If we wish to calculate the Green's function on the LHS of Eq.~(\ref{eq:greensrelation}), we may choose $l^*$ such that $e^{l^*}=(E_0/E)$ where $E_0$ is 
an energy comparable to the bandwidth (i.e. perturbation theory should be valid there). 

Eq.~(\ref{eq:greensrelation}) is also useful when $E$ is large and the Born approximation is valid. In this case, it leads directly to
a scaling relation for $1/\tau$. It is
\begin{eqnarray}
\frac{1}{\tau(bE, \{g(dl)\})} = \frac{b}{\tau(E, \{g\})},
\label{eq:tauscale}
\end{eqnarray}
where contributions of $O(g^2)$ have been ignored.
Since resistivity is proportional to $1/\tau$, Eq.~(\ref{eq:tauscale}) may be used to calculate its temperature or frequency dependence in regimes
where perturbation theory is valid (i.e. well above spin and charge gaps). In order to do this, the scaling behaviour of couplings which lead to scattering
in the $\rho^+$ channel must be known. This information is easily determined in the Boson formalism (see Ref.~\cite{vond}). As an example, the disorder
strengths, $D_{a/b}$, in the Hamiltonian of Eq.~(\ref{eq:spingapham}) scale as $D^\prime_{a/b}=D_{a/b}b^{3-(K_{\rho^+}-K_{\rho^-})/2}$. 
This yields a high temperature resistivity of the form
\begin{eqnarray}
\rho_c\propto a_1D_aT^{\frac{K_{\rho^+}}{2}+\frac{K_{\rho^-}}{2}-2}+a_3D_bT^{\frac{K_{\rho^+}}{2}+\frac{K_{\rho^-}}{2}-2}\nonumber\\
\label{eq:gapres}
\end{eqnarray}
where the $a_i$'s are constants. This form for resistivity is valid above the $\rho^-$ gap if spin fluctuations are sufficiently suppressed. 
In the next section, we argue that the Luttinger parameters cannot deviate significantly from unity 
without moving the spin and/or charge gaps to extremely high or low energies. Therefore, Eq.~(\ref{eq:gapres}) implies that $\rho_c$ will have a monotonically
decreasing temperature dependence in the presence of pairing fluctuations and disorder. This is inconsistent with the linear power law regime observed
in $\mathrm{Sr_{3}Ca_{11}Cu_{24}O_{41}}$.

Although some earlier experimental evidence points to a $300$ K spin gap for the strontium cuprate ladder, NMR results from Fujiyama et al.\cite{fuj} indicate the existence of spin fluctuations
at temperatures as low as $60$ K - about the same temperature at which charge localization occurs. The Hamiltonian consisting of Eq.~(\ref{eq:newform}) and~(\ref{eq:bosdis}), which
allows for spin fluctuations, may, therefore, be a more appropriate model in the power law regime. Performing the same scaling analysis which led to Eq.~(\ref{eq:gapres}) on this model yields
a high temperature resistivity of the form
\begin{eqnarray}
\rho_{c}&\propto&a_1D_1\left(\frac{a_2}{T}\right)^{2-\frac{1}{2}\left(K_{\rho^+}+K_{\rho^-}+K_{\sigma^+}+K_{\sigma^-}\right)}\nonumber\\
&+&a_3D_2\left(\frac{a_4}{T}\right)^{2-\frac{1}{2}\left(K_{\rho^+}+K_{\rho^-}+K_{\sigma^+}+K_{\sigma^-}\right)}\nonumber\\
&+&a_5D_3\left(\frac{a_6}{T}\right)^{2-\frac{1}{2}\left(K_{\rho^+}+K_{\rho^-}+K_{\sigma^+}+\frac{1}{K_{\sigma^-}}\right)}\nonumber\\
&+&a_7D_4\left(\frac{a_8}{T}\right)^{2-\frac{1}{2}\left(K_{\rho^+}+K_{\rho^-}+K_{\sigma^-}+\frac{1}{K_{\sigma^-}}\right)}.
\label{eq:secondrhodc}
\end{eqnarray}
For Luttinger parameters close to unity, this resistivity will have little temperature dependence.

Turning to umklapp processes, if the system is not far from a special filling (i.e. one of the $\delta$'s in Eq.~(\ref{eq:umklapp}) is small), then, at 
high temperatures, Eq.~(\ref{eq:tauscale}) can be used to determine $\rho_c$ as before. For the case of $\delta_{b+a}$ small, it is
\begin{eqnarray}
\rho_{c}&\propto&a_1g_{u1}^2\left(\frac{a_2}{T}\right)^{3-2\left(K_{\rho^+}+\frac{1}{K_{\rho^-}}\right)}\nonumber\\
&+&a_3g_{u2}^2\left(\frac{a_4}{T}\right)^{3-2\left(K_{\rho^+}+K_{\sigma^-}\right)}\nonumber\\
&+&a_5g_{u3}^2\left(\frac{a_6}{T}\right)^{3-2\left(K_{\rho^+}+\frac{1}{K_{\sigma^-}}\right)}.
\label{eq:umrhodc}
\end{eqnarray}
If the Luttinger parameters are close to unity, then this resistivity is approximately linear in $T$. 
For the case of $\delta_{b/a}$ small, similar results hold. 

At this point, an important difference between the effect of umklapp processes on the conductivity of single chains and two-leg ladders becomes apparent.
In the case of a single chain, if $\delta\ll T$ but not zero ($\delta = 0$ gives rise to a charge gap) and $K_\rho\approx 1$, then umklapp processes lead to a linear $T$ dependence
for resistivity regardless of the behaviour of the spin mode (which is completely decoupled from the charge mode)~\cite{gia2}. However, in the case of the two-leg ladder,
charge and spin gaps can alter this behaviour. For example, if a specific ladder material has a large spin gap, then a regime in which fluctuations of the $\rho^-$
mode are allowed but spin fluctuations are suppressed might exist. For $\delta_{b+a}$ small, an effective Hamiltonian for this regime may be derived from
Eq.~(\ref{eq:umklapp}) by integrating over fluctuations in the $\theta_{\sigma^-}$ field. The result would be terms in the DC resistivity which are proportional
to $T^5$, $T$ and $1/T$. Below the $\rho^-$ gap, all fillings which give rise to significant umklapp scattering would give $T^5$ and $1/T$ contributions~\cite{pat}.

\section{\label{sec:RG}Renormalization Group Analysis}

In the previous section it was demonstrated that, at high energies, umklapp processes can contribute a linear temperature dependence to the DC resistivity
of two-leg ladder compounds. In this section and the next, we discuss what happens at lower energy and attempt to explain the conductivity data of Osafune et al.~\cite{osa}
Based on the discussion of the previous sections, we take our model to consist of the clean Hamiltonian of Eq.~(\ref{eq:ham1}), the $\delta_{b+a}$ term of the umklapp Hamiltonian of
Eq.~(\ref{eq:umklapp}) (we drop the $b+a$ subscript from here on), and the disorder Hamiltonian,
\begin{eqnarray}
H_{\mathrm{dis}}&=&\int \frac{dx}{a^2}\, \xi e^{-\mathrm{i}\sqrt{4\pi}\prp}+\xi^* e^{\mathrm{i}\sqrt{4\pi}\prp},
\label{eq:genedis}
\end{eqnarray}
used in Ref.\cite{kim}. Note that $H_{\mathrm{dis}}$ can be derived from Eq.~(\ref{eq:bosdis}) by integrating over fluctuations in the $\sigma^{\pm}$ and the $\rho^-$ modes which exist
when $g_{b/as}$ and $g_c$ are large. Since these couplings are strongly relevant, Eq.~(\ref{eq:genedis}) is the correct disorder Hamiltonian at low energies. 
Here, we assume that the disorder couplings appearing in Eq.~(\ref{eq:genedis}) are sufficiently weak that $H_{\mathrm{dis}}$ has little effect on transport properties
except at these low energies.

RG equations may be derived for our model by scaling space-time variables while adjusting coupling constants to maintain the real space cutoff, $a$. Following
standard methods~\cite{car}, this procedure has been carried out to second order in the couplings, with the results shown in Appendix~\ref{ap:RG}. These equations have a number of important implications.
In weak coupling, the $\sigma^{\pm}$ and the $\rho^-$ modes are gapped due to the relevancy of $g_1$, $g_2$, $g_{as}$ and $g_{bs}$. Numerical work~\cite{noa} has shown that these
gaps persist in the strong coupling case. Therefore, the parameters of our phenomenological Hamiltonian must have values such that the couplings which generate the gaps grow large 
at the appropriate energy scale. From the RG equations
it is also clear that, when the interaction couplings are small, the energy scale of the gaps is determined by the Luttinger parameters. For the case of a one-dimensional chain, there is only 
one Luttinger parameter, $K_{\sigma}$, relating to spin. This parameter is restricted to values near unity~\cite{fra}. We will assume that it is reasonable to take $K_{\sigma^\pm}$ near unity as well. 
In fact, if either of $K_{\sigma^{\pm}}$ or $K_{\rho^{-}}$ deviates significantly from unity, then one or more gaps will open at very high energy (near $E_0$) or at very low energy. 
This would be inconsistent with a gap magnitude of approximately $60$ K.
Taken together with the fact that $K_{\rho^+}<1$ for repulsive density interactions, these Luttinger parameter restrictions indicate that umklapp
scattering gives rise to a nearly linear $T$ dependence for resistivity at high energies (see Eq.~(\ref{eq:umrhodc})). In addition, numerical integration of the RG equations
shows that the Luttinger parameters do not undergo significant renormalization until the gap energy is reached. Therefore, this linear $T$ behaviour
should persist until just above the gap energies, well below $E_0$.

Given that scaling arguments strongly suggest umklapp processes as the mechanism responsible for the power law resistivity feature of $\mathrm{Sr_{3}Ca_{11}Cu_{24}O_{41}}$, it would be useful to
perform conductivity calculations which are valid at all energy scales. To accomplish this, perturbative calculations along with the RG equations and Eq.~(\ref{eq:greenscale}) may 
be used. A number of simplifications can be made at this point. First, from numerical integration, we find that the Luttinger parameters and the
velocities flow slowly at high energies (their flow rate is proportional to $O(g^2)$). At low energies, renormalization due to umklapp processes is attenuated by a rapid growth
of $\delta$. Therefore, we ignore renormalization of Luttinger parameters and velocities due to umklapp processes. Within this approximation, the velocities will all flow
to the same value until localization takes over at low energy. This implies that, above the localization energy, there is no qualitative difference between 
conductivities calculated with different velocity parameters. Taking all velocities as equal also allows perturbative conductivity calculations to be performed in a 
straightforward manner using standard results~\cite{shu2,bou}. Finally, all gap
generating couplings will flow slowly under renormalization until they become $O(1)$, at which point they diverge rapidly. The Luttinger parameters also flow slowly until
this point. Given that the $\mathrm{Sr_{14-x}Ca_xCu_{24}O_{41}}$ ladder has spin fluctuations at energies as low as the charge localization energy, we ignore renormalization of the Luttinger parameters.
This is qualitatively valid since these parameters only flow in a significant way after wavefunction renormalization becomes the dominant factor in scaling (see Eq.~(\ref{eq:greenscale})).

\section{\label{sec:pert}Perturbative Calculation of the Memory Function}

In order to make use of Eq.~(\ref{eq:greenscale}), perturbative calculations of $\rho_c$ and $\sigma_c$ are required. Following Giamarchi~\cite{gia2}, the memory function formalism~\cite{got}
may be implemented for this purpose. We proceed by writing the conductivity as
\begin{eqnarray}
\sigma = \mathrm{i}Q\frac{1}{\omega+M(\omega)}
\label{eq:memcond}
\end{eqnarray}
where $Q$ is the response of the free electron system and $M$ is the memory function given by
\begin{eqnarray}
M(\omega)=\frac{\left<F;F\right>_0(\omega)-\left<F;F\right>_0(0)}{\omega Q},
\label{eq:memfun}
\end{eqnarray}
with
\begin{eqnarray}
\left<F;F\right>_0=-\mathrm{i}\int_{-\infty}^{\infty} dx\int_0^{\infty} dt\, e^{-\mathrm{i}\omega t}\left<\left[F(x,t),F(0)\right]\right>_0\nonumber\\
\end{eqnarray}
and $F=\left[j,H\right]$.
From Eq.~(\ref{eq:umklapp}),~(\ref{eq:currentop}) and~(\ref{eq:genedis}) we find
\begin{eqnarray}
F&=&\frac{\mathrm{i}4uK_{\rho^+}}{a^2}\left[\sin\left(\sqrt{4\pi}\prp+\delta x\right)\left\{g_{u1}\cos\sqrt{4\pi}\trm\right.\right.\nonumber\\
&-&\left.\left. g_{u2}\cos\sqrt{4\pi}\psm-g_{u3}\cos\sqrt{4\pi}\tsm\right\}\right.\nonumber\\
&+&\left.\xi^*(x)e^{\mathrm{i}\sqrt{4\pi}\prp}-\xi(x)e^{-\mathrm{i}\sqrt{4\pi}\prp}\right]
\end{eqnarray}
where only the $\delta_{b+a}$ term from the umklapp Hamiltonian has been used and all velocities have been taken as equal. Using the results of Schulz~\cite{shu2}, and Schulz and Bourbonnais~\cite{bou},
$\left<F;F\right>_0(\omega)$ may be calculated in a straightforward manner. The memory function of Eq.~(\ref{eq:memfun}) then becomes $M=M_\mathrm{um}+M_\mathrm{dis}$, with
\begin{eqnarray}
M_\mathrm{um}&=&\frac{\pi K_{\rho^+}}{a^2}\frac{1}{\omega}\sum_{\chi=1}^3g_{u\chi}^2\sin\left(\pi\kappa_\chi\right)
\left(\frac{2\pi aT}{u}\right)^{2\left(\kappa_\chi-1\right)}\nonumber\\
&\times&\left(\mathcal{B}_\chi^\omega (\delta)-\mathcal{B}_\chi^0 (\delta)\right)
\end{eqnarray}
and
\begin{eqnarray}
M_\mathrm{dis}&=&4D\sin(\pi K_{\rho^+})\left(\frac{2\pi a T}{u}\right)^{2(K_{\rho^+}-1)}\nonumber\\
&\times&\int_{-\pi}^\pi dp\, \left(\mathcal{B}_{\rho^+}^\omega (\frac{p}{a})-\mathcal{B}_{\rho^+}^0 (\frac{p}{a})\right)
\end{eqnarray}
where
\begin{eqnarray}
\mathcal{B}_\chi^\omega (\delta)&=&B\left(\frac{\kappa_\chi}{2}-\mathrm{i}S_+^\omega(\delta),1-\kappa_\chi\right)\nonumber\\
&\times&B\left(\frac{\kappa_\chi}{2}-\mathrm{i}S_-^\omega(\delta),1-\kappa_\chi\right), \nonumber\\
\mathcal{B}_{\rho^+}^\omega (\frac{p}{a})&=&B\left(\frac{K_{\rho^+}}{2}-\mathrm{i}S_+^\omega(\frac{p}{a}),1-K_{\rho^+}\right)\nonumber\\
&\times&B\left(\frac{K_{\rho^+}}{2}-\mathrm{i}S_-^\omega(\frac{p}{a}),1-K_{\rho^+}\right),\nonumber\\
S_\pm^\omega(x)&=&\frac{\omega\pm ux}{4\pi T},
\end{eqnarray}
and
\begin{eqnarray}
\kappa_\chi=\left\{\begin{array}{r@{\quad:\quad}l}
K_{\rho^+}+\frac{1}{K_{\rho^-}}& \chi=1\\
K_{\rho^+}+K_{\sigma^-}&\chi=2\\
K_{\rho^+}+\frac{1}{K_{\sigma^-}}&\chi=3
\end{array}\right.
\end{eqnarray}
Here $B(x,y)$ is the Beta function and $Q$ has been taken as $4uK_{\rho^+}/\pi$, twice the value for a single band model~\cite{sha}. Also,
the $p$ integral in $M_\mathrm{dis}$ arises because the disorder field, $\xi$, is Fourier transformed before disorder averaging is performed.

In this work, the zero frequency and zero temperature limits of the memory function will be of interest. For $T\rightarrow 0$, $\omega\ne 0$ we have
\begin{multline}
M_\mathrm{um}\rightarrow\\
\frac{\pi K_{\rho^+}}{a^2}\frac{1}{\omega}\sum_{\chi=1}^3g_{u\chi}^2\Gamma^2(1-\kappa_\chi)\sin(\pi\kappa_\chi)\left(\frac{a}{2u}\right)^{2(\kappa_\chi-1)}\\
\times\left\{\overline{\left[\left(u\delta\right)^2-\omega^2\right]}^{\kappa_\chi-1}-\left(u\delta\right)^{2(\kappa_\chi-1)}\right\}.
\end{multline}
and
\begin{multline}
M_\mathrm{dis}\rightarrow\\ 
\frac{4DK_{\rho^+}}{a^2}\frac{1}{\omega}\sin(\pi K_{\rho^+})\Gamma^2(1-K_{\rho^+})\left(\frac{1}{2u}\right)^{2(K_{\rho^+}-1)}\\
\times\int_{-\pi}^\pi dp\, \left\{\overline{\left[\left(up\right)^2-\omega^2\right]}^{K_{\rho^+}-1}-\left(up\right)^{2(K_{\rho^+}-1)}\right\}
\end{multline}
For the case $\omega \rightarrow 0$, $T\ne 0$ we have
\begin{eqnarray}
M_\mathrm{um}&\rightarrow&\frac{\mathrm{i}\pi K_{\rho^+}}{a^2}\sum_{\chi=1}^3g_{u\chi}^2\Gamma^2(1-\kappa_\chi)\sin(\pi\kappa_\chi)\nonumber\\
&\times&\left(\frac{2\pi aT}{u}\right)^{2(\kappa_\chi-1)}\left(\frac{1}{4\pi T}\right)\frac{\Gamma(z_\chi)\Gamma(\bar{z}_\chi)}{\Gamma(1-z_\chi)\Gamma(1-\bar{z}_\chi)}\nonumber\\
&\times&\left[\Psi(1-z_\chi)+\Psi(1-\bar{z}_\chi)-\Psi(z_\chi)-\Psi(\bar{z}_\chi)\right]
\end{eqnarray}
and
\begin{eqnarray}
M_\mathrm{dis}&\rightarrow&\frac{\mathrm{i}4DK_{\rho^+}}{a^2}\Gamma^2(1-K_{\rho^+})\sin(\pi K_{\rho^+})\left(\frac{1}{4\pi T}\right)\nonumber\\
&\times&\left(\frac{2\pi aT}{u}\right)^{2(K_{\rho^+}-1)}\int_{-\pi}^\pi dp\, \frac{\Gamma(y_p)\Gamma(\bar{y}_p)}{\Gamma(1-y_p)\Gamma(1-\bar{y}_p)}\nonumber\\
&\times&\left[\Psi(1-y_p)+\Psi(1-\bar{y}_p)-\Psi(y_p)-\Psi(\bar{y}_p)\right]
\end{eqnarray}
where $\Psi(x)=\Gamma^\prime (x)/\Gamma(x)$,
\begin{eqnarray}
z_\chi&=&\frac{\kappa_\chi}{2}+\mathrm{i}\frac{u\delta}{4\pi T}
\end{eqnarray}
and
\begin{eqnarray}
y_p&=&\frac{K_{\rho^+}}{2}+\mathrm{i}\frac{up}{4\pi aT}.
\end{eqnarray}

\section{\label{sec:curves}Results From Scaling}

Using the memory functions from the previous section along with Eq.~(\ref{eq:realcond}),~(\ref{eq:greenscale}) and~(\ref{eq:RGequations}), conductivity and resistivity 
curves may be produced for a given set of couplings. To proceed, it is useful to rewrite Eq.~(\ref{eq:memcond}) using Eq.~(\ref{eq:realcond}) and~(\ref{eq:greenscale}) to obtain
a scaling relation for resistivity/conductivity in terms of the memory function. For resistivity we have 
\begin{eqnarray}
\rho(T,\{g\})=-\frac{\mathrm{i}\pi e^{-l^*}}{4uK_{\rho^+}(l^*)}e^{\int_0^{l^*}dl\, \gamma(l)}M(e^{l^*}T,\{g(l^*)\})\nonumber\\
\label{eq:memresscale}
\end{eqnarray}
and for conductivity,
\begin{eqnarray}
\Re\sigma(\omega,\{g\})&=&-\frac{4u}{\pi\omega}e^{-\int_0^{l^*}dl\, \gamma(l)}\nonumber\\
&\times&\Im\left[\frac{e^{l^*}\omega K_{\rho^+}(l^*)}{e^{l^*}\omega
+M\left(e^{l^*}\omega,\{g(l^*)\}\right)}\right].\nonumber\\
\label{eq:memcondscale}
\end{eqnarray}

In Sec.~\ref{sec:RG}, it was argued that the Luttinger parameters and velocities do not renormalize significantly until the charge localization energy is reached. Therefore,
in using Eq.~(\ref{eq:memresscale}) and~(\ref{eq:memcondscale}), renormalization of these quantities will be ignored. Within this approximation, the parameters which affect
the resistivity/conductivity curves are $D, g_{ui}, u, \delta, K_{\rho^+}, K_{\rho^-} \mbox{ and } K_{\sigma^-}$. Given any set of these parameters, we can now generate such curves.
To set units, we make the transformation
\begin{align}
T&\rightarrow k_BT, & \omega&\rightarrow \hbar\omega,\nonumber\\
g_i&\rightarrow g_ita, & u&\rightarrow uta,\nonumber\\
D&\rightarrow Dt^2a^2, & \rho&\rightarrow\frac{1}{e^2}\rho,\nonumber\\
\sigma&\rightarrow e^2\sigma
\end{align}
where $e$ is the electron charge. 

\begin{figure}
\resizebox{3in}{2.6in}{\includegraphics{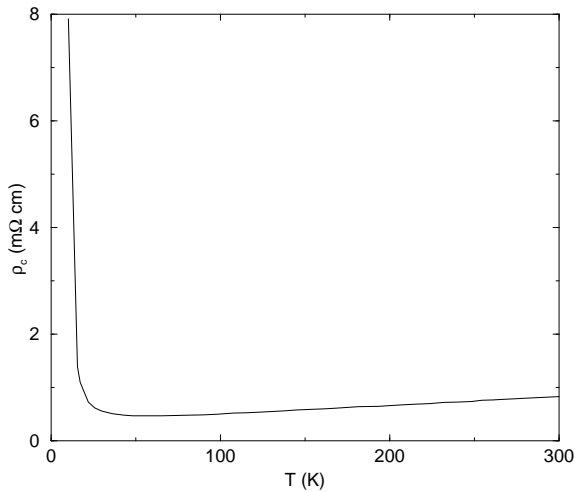}}
\caption{\label{fig:res}$\rho_c(T)$ vs. $T$.}
\end{figure}

By trial and error, it is simple to choose parameters which generate curves that resemble those measured by Osafune et al.~\cite{osa} We first recall that the Luttinger parameters should not be
far from one, $E_0$ must be well above the highest temperature or wavenumber of interest, $K_{\rho^+}$ must be less than one and $\delta$ should be small. It is also of
interest to note that, quantitatively, the resistivity and conductivity curves are most sensitive to changes in $D$ and $K_{\rho^+}$.
For both resistivity and conductivity,
we choose  $E_0=0.25$ eV, $ut=1.1$ eV, $K_{\rho^+} = 0.9$, $K_{\rho^-}=1.15$, $K_{\sigma^-}=0.85$, $g_{u1}t=0.4$ eV, $g_{u2}t=0$ eV, $g_{u3}t= 0.5$ eV, $\delta=0.022$ and $D=0.00034$ eV$^2$.
The resulting resistivity curve, shown in Fig.~\ref{fig:res}, demonstrates insulating behavior beginning below $60$ K and a monotonically increasing power law regime above that temperature.
The localization temperature is mainly determined by $K_{\rho^+}$, the exact nature of the power law above $60$ K is determined by the Luttinger parameters, the separation between the two
regimes is determined by $\delta$ and the relative heights of the various parts of the curve are determined by $\{g_{ui}\}$ and $D$. The velocity, $u$, plays a role in
determining the overall scale of the curve, as well as how fast the resistivity diverges at low $T$.
\begin{figure}
\resizebox{3in}{2.6in}{\includegraphics{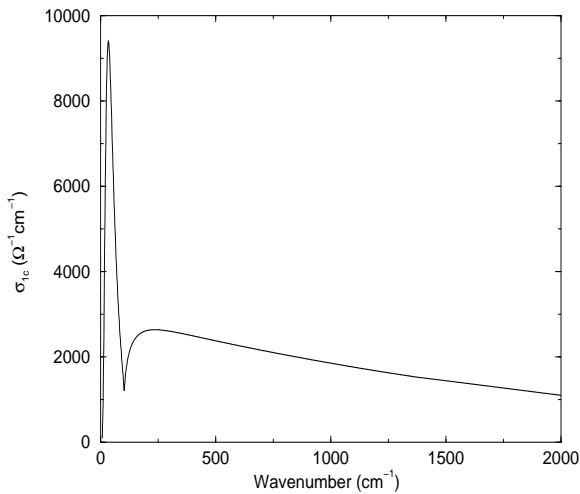}}
\caption{\label{fig:op}$\sigma_{1c}(\omega)$ vs. wavenumber.}
\end{figure}
Our conductivity curve, shown in Fig.~\ref{fig:op}, demonstrates a finite frequency, Drude-like peak with additional spectral weight at higher wavenumbers. The position of the peak
is determined mainly by $K_{\rho^+}$, the peak height and attenuation are determined by $u$ and $D$, the relative height of the spectral weight beyond the peak is determined
by $\{g_{ui}\}$, the nature of the power law at high wavenumbers is determined by the Luttinger parameters, and the wavenumber at which absorption due to umklapp processes begins 
is determined by $\delta$.

Although we have not done a detailed numerical fit to the data, our resistivity and conductivity curves show good qualitative agreement with the 
measurements of Osafune et al.~\cite{osa}. However, it it also clear that one can achieve a better quantitative fit to the data by choosing different values of $D$ and $K_{\rho^+}$ to independently 
fit the resistivity and conductivity curves. This quantitative discrepancy is likely due to the fact that the renormalization of the Luttinger parameters and velocities was ignored.

\section{\label{sec:conc}Conclusions}

In this paper, we have used RG and scaling techniques to argue that the conductivity properties of the two-leg ladder compound $\mathrm{Sr_{14-x}Ca_xCu_{24}O_{41}}$ can
be explained by a combination of disorder and umklapp effects. It has been shown that quenched disorder, whether in the presence of pairing fluctuations alone or in the
presence of pairing and spin fluctuations, does not contribute a monotonically increasing temperature dependence to the resistivity at high $T$. However, if spin and pairing
fluctuations persist down to the charge localization temperature, then umklapp process can contribute an approximately linear temperature dependence if the system
is near an appropriate filling. Experimental evidence from Fujiyama et al.~\cite{fuj} suggests that such fluctuations do exist.

\appendix
\section{\label{ap:RG}RG Equations}

We have derived RG equations for the model described in Sec.~\ref{sec:RG} using standard techniques. To second order in the interaction couplings they are
\begin{align}
\frac{dg_1}{dl}& = g_1\left(2-\frac{1}{K_{\rho^-}}-K_{\sigma^+}\right),\nonumber\\
\frac{dg_2}{dl}& = g_2\left(2-\frac{1}{K_{\rho^-}}-K_{\sigma^-}\right),\nonumber\\
\frac{dg_{bs}}{dl}& = g_{bs}\left(2-K_{\sigma^+}-K_{\sigma^-}\right),\nonumber\\
\frac{dg_{as}}{dl}& = g_{as}\left(2-K_{\sigma^+}-K_{\sigma^-}\right),\nonumber\\
\frac{dg_{u1}}{dl}& = g_{u1}\left(2-K_{\rho^+}-\frac{1}{K_{\rho^-}}\right),\nonumber\\
\frac{dg_{u2}}{dl}& = g_{u2}\left(2-K_{\rho^+}-K_{\sigma^-}\right),\nonumber\\
\frac{dg_{u3}}{dl}& = g_{u3}\left(2-K_{\rho^+}-\frac{1}{K_{\sigma^-}}\right),\nonumber
\end{align}
\vspace{-0.22in}
\begin{eqnarray}
\frac{d\delta}{dl} = \delta, \frac{dD}{dl} = D(3-2K_{\rho^+}),  \nonumber
\end{eqnarray}
\vspace{-0.3in}
\begin{widetext}
{\allowdisplaybreaks
\begin{eqnarray}
\frac{du_{\rho^+}}{dl}&=& \frac{(2\pi)^2(g_{u1}^2+g_{u2}^2+g_{u3}^2)K_{\rho^+}}{4u_0}\left[\left(1-\frac{u_{\rho^+}^2}{u_0^2}\right)
J_0(\delta a)-\left(1+\frac{u_{\rho^+}^2}{u_0^2}\right)J_2(\delta a)\right]-\frac{8\pi DK_{\rho^+}}{u_{0}},\nonumber\\
\frac{du_{\rho^-}}{dl}&=& \frac{(2\pi)^2}{4u_0}\frac{1}{K_{\rho^-}}\left[\left(1-\frac{u_{\rho^-}^2}{u_0^2}\right)
\left(\frac{g_1^2+g_2^2+2J_0(\delta a)g_{u1}^2}{2}\right)-\left(1+\frac{u_{\rho^-}^2}{u_0^2}\right)J_2(\delta a)g_{u1}^2\right],\nonumber\\
\frac{du_{\sigma^+}}{dl}&=& \frac{(2\pi)^2}{8u_0}\left(1-\frac{u_{\sigma^+}^2}{u_0^2}\right)\left(g_1^2+g_+^2+g_-^2\right)K_{\sigma^+},\nonumber\\
\frac{du_{\sigma^-}}{dl}&=& \frac{(2\pi)^2}{8u_0}\left[\left(1-\frac{u_{\sigma^-}^2}{u_0^2}\right)\biggl\{\left(g_2^2+g_+^2+g_-^2+2J_0(\delta a)g_{u2}^2
\right)K_{\sigma^-}+\frac{2J_0(\delta a)g_{u3}^2}{K_{\sigma^-}}\biggr\}\right.\nonumber\\
&-&\left.\left(1+\frac{u_{\sigma^-}^2}{u_0^2}\right)\left\{2J_2(\delta a)g_{u2}^2K_{\sigma^-}
+\frac{2J_2(\delta a)g_{u3}^2}{K_{\sigma^-}}\right\}\right],\nonumber\\
\frac{dK_{\rho^+}}{dl}&=& \frac{(2\pi)^2(g_{u1}^2+g_{u2}^2+g_{u3}^2)K_{\rho^+}^2}{4u_{\rho^+}u_0}\left[-\left(1+\frac{u_{\rho^+}^2}{u_0^2}\right)
J_0(\delta a)+\left(1-\frac{u_{\rho^+}^2}{u_0^2}\right)J_2(\delta a)\right]-\frac{8\pi DK_{\rho^+}^2}{u_{\rho^+}u_0},\nonumber\\
\frac{dK_{\rho^-}}{dl}&=& \frac{(2\pi)^2}{4u_{\rho^-}u_0}\left[\left(1+\frac{u_{\rho^-}^2}{u_0^2}\right)
\left(\frac{g_1^2+g_2^2+2J_0(\delta a)g_{u1}^2}{2}\right)-\left(1-\frac{u_{\rho^-}^2}{u_0^2}\right)J_2(\delta a)g_{u1}^2\right],\nonumber\\
\frac{dK_{\sigma^+}}{dl}&=& -\frac{(2\pi)^2}{8u_{\sigma^+}u_0}\left(1+\frac{u_{\sigma^+}^2}{u_0^2}\right)\left(g_1^2+g_+^2+g_-^2\right)K_{\sigma^+}^2,\nonumber\\
\frac{dK_{\sigma^-}}{dl}&=& \frac{(2\pi)^2}{8u_{\sigma^-}u_0}\left[-\left(1+\frac{u_{\sigma^-}^2}{u_0^2}\right)\biggl\{\left(g_2^2+g_+^2+g_-^2+2J_0(\delta a)g_{u2}^2
\right)K_{\sigma^-}^2-2J_0(\delta a)g_{u3}^2\biggr\}\right.\nonumber\\
&+&\left.\left(1-\frac{u_{\sigma^-}^2}{u_0^2}\right)\biggl\{2J_2(\delta a)g_{u2}^2K_{\sigma^-}^2-2J_2(\delta a)g_{u3}^2\biggr\}\right]
\label{eq:RGequations}
\end{eqnarray} }
\end{widetext}
where $J_{0/2}$ are the zeroth/second order Bessel functions, the $b+a$ subscript has been dropped from $\delta_{b+a}$, and $u_0$ was introduced to set the units of the time
cutoff correctly. It is immediately clear that the velocities and Luttinger parameters initially flow slowly since their rate of change is $O(g^2)$. Also, $\delta$ grows exponentially
so that the Bessel functions which appear in the equations for the velocities and Luttinger parameters go to zero at low energies. The contribution of umklapp processes to the renormalization
of velocities and Luttinger parameters is neglected in this work by dropping all terms involving Bessel functions. Within this approximation, the velocities other than $u_{\rho^+}$ 
will all flow to $u_0$ while $u_{\rho^+}$ will only flow in a significant way when localization takes over at low energies. At this point, wavefunction renormalization
becomes the dominant effect so we may ignore renormalization of the velocities as well. Finally, the feedback between the Luttinger parameters and the couplings
which renormalize it is such that the Luttinger parameters flow slowly until couplings become $O(1)$. At this point, both couplings and Luttinger parameters diverge rapidly.
However, if the gaps in the system occur below the charge localization energy, then this won't happen until wavefunction renormalization becomes the dominant effect.
Under these circumstances, neglecting Luttinger parameter renormalization should not introduce any qualitative errors in calculations of conductivity (although quantitatively
there will likely be some effect).

\begin{acknowledgments}
The authors wish to thank A.J. Berlinsky and C. Bourbonnais for valuable discussions. 
In addition, we acknowledge NSERC for financial support.
\end{acknowledgments}

\end{document}